\documentclass[aps,prl,reprint,superscriptaddress]{revtex4-1}
\usepackage{graphicx}
\usepackage[rightcaption]{sidecap}
\usepackage{braket}
\usepackage[utf8]{inputenc}
\usepackage{hyperref}
\usepackage{amsmath} 
\usepackage{amssymb}  
\usepackage{amstext}  
\usepackage{graphicx}
\usepackage{amssymb}
\usepackage{color}

\newcommand{\yb}[1]{$^{#1}$Yb$^+$}
\newcommand{\state}[3]{$^{#1}${#2}$_{#3}$}

\begin{document}
\title{Direct photonic coupling of a semiconductor quantum dot and a trapped ion}

\author{H. M. Meyer}
\affiliation{Cavendish Laboratory, University of Cambridge, JJ Thomson Avenue, Cambridge CB3 0HE, United Kingdom}
\affiliation{Physikalisches Institut, University of Bonn, Wegelerstrasse 8, 53115 Bonn, Germany}

\author{R. Stockill}
\affiliation{Cavendish Laboratory, University of Cambridge, JJ Thomson Avenue, Cambridge CB3 0HE, United Kingdom}

\author{M. Steiner}
\altaffiliation[present address:]{Centre for Quantum Technologies, National University of Singapore, 3 Science Drive 2; Singapore 117543, Singapore}
\affiliation{Cavendish Laboratory, University of Cambridge, JJ Thomson Avenue, Cambridge CB3 0HE, United Kingdom}

\author{C. Le Gall}
\affiliation{Cavendish Laboratory, University of Cambridge, JJ Thomson Avenue, Cambridge CB3 0HE, United Kingdom}

\author{C. Matthiesen}
\affiliation{Cavendish Laboratory, University of Cambridge, JJ Thomson Avenue, Cambridge CB3 0HE, United Kingdom}

\author{E. Clarke}
\affiliation{EPSRC National Centre for III-V Technologies, University of Sheffield, Sheffield, S1 3JD, UK}

\author{ A. Ludwig}
\affiliation{Lehrstuhl f\"ur Angewandte Festk\"orperphysik, Ruhr-Universit\"at, 44780 Bochum, Germany}

\author{J. Reichel}
\affiliation{Laboratoire Kastler Brossel, \'Ecole Normale Sup\'erieure, 24 Rue Lhomond, 75005 Paris, France}

\author{M. Atat{\"u}re}
\email[Electronic address: ]{ma424@cam.ac.uk}
\affiliation{Cavendish Laboratory, University of Cambridge, JJ Thomson Avenue, Cambridge CB3 0HE, United Kingdom}

\author{M. K{\"o}hl}
\email[Electronic address: ]{michael.koehl@uni-bonn.de}
\affiliation{Cavendish Laboratory, University of Cambridge, JJ Thomson Avenue, Cambridge CB3 0HE, United Kingdom}
\affiliation{Physikalisches Institut, University of Bonn, Wegelerstrasse 8, 53115 Bonn, Germany}

\begin{abstract}
Coupling individual quantum systems lies at the heart of building scalable quantum networks. 
Here, we report the first direct photonic coupling between a semiconductor quantum dot and a trapped ion and we demonstrate that single photons generated by a quantum dot controllably change the internal state of an $\textrm{Yb}^+$ ion. We ameliorate the effect of the sixty-fold mismatch of the radiative linewidths with coherent photon generation and a high-finesse fiber-based optical cavity enhancing the coupling between the single photon and the ion. The transfer of information presented here via the classical correlations between the $\sigma_z$-projection of the quantum-dot spin and the internal state of the ion provides a promising step towards quantum state-transfer in a hybrid photonic network.
\end{abstract}

\pacs{
37.30.+i, 
03.67.-a,
42.50.Pq
}

\date{\today}

\maketitle

Single atoms and ions are among the key players in the realization of elementary quantum information processing protocols \cite{nielsen2000}. High-fidelity state preparation and readout paired with long coherence times of internal and external degrees of freedom have enabled the implementation of small quantum processing units \cite{Kimble2008,BlattWineland2008,Monroe2013}. In recent years, optically active spin qubits in the solid state \cite{Awschalom2013}, such as semiconductor quantum dots (QDs) \cite{DeGreve2013} and impurity centers in diamond \cite{Childress2013}, have emerged as complementary systems. Albeit having shorter coherence times, these systems offer ultrafast quantum control via larger electrical dipole moments \cite{Press2008} and on-chip integration  \cite{Englund2007,Laucht2012,Hausmann2012,Luxmoore2013,Arcari2014} without the need for a continuously operating trap architecture. While prototype photonic networks  of identical constituents, such as single atoms \cite{Moehring2007,Ritter2012}, or spins in diamond \cite{Dutt2007, Pfaff2014} have been demonstrated, the concept of hybrid quantum networks has recently been proposed as an exciting alternative \cite{Tian2004,Wallquist2009,Waks2009}. Initial progress towards hybrid systems includes single-photon sources coupled to atomic vapours  \cite{Akopian2011, Siyushev2014} and superconducting qubits to solid-state spin ensembles  \cite{Zhu2011, Kubo2011}. These specific experiments rely on large ensembles and, in  some cases, spatial proximity to within the coherence length of the interaction. The formation of a modular network where fundamentally differing individual quantum systems communicate over long distances is an important goal, which so far has remained elusive.\\
\begin{figure*}
\includegraphics[width=1\textwidth,angle=0]{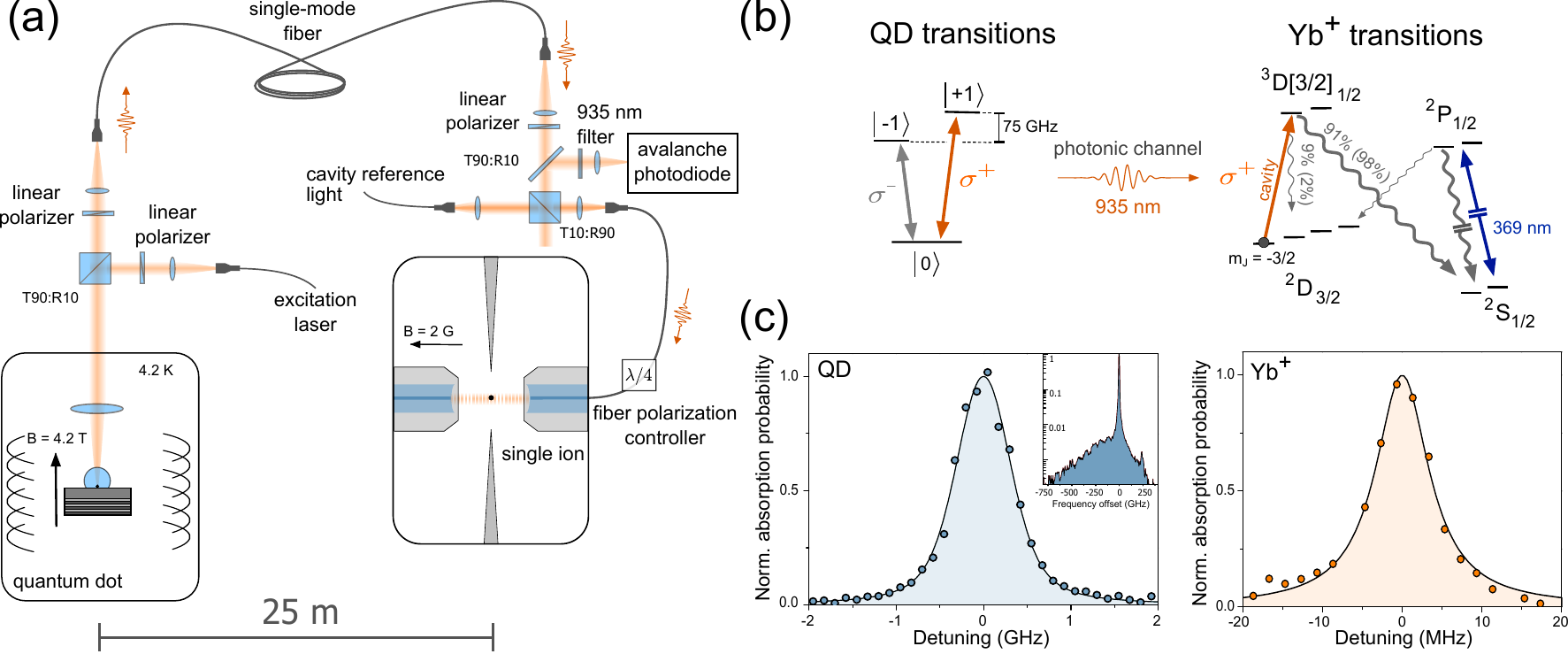}
\caption{ \label{fig1} (Color online) Experimental setup and optical transitions of the atomic and solid-state nodes. (a) The QD is located inside a magneto-optical cryostat operating at $4.2$ K and $4.2$ T. Single photons generated resonantly at 935\,nm are sent to the atomic node via a 50\,m single-mode fiber. The ion is placed inside a high-finesse optical cavity resonant with the ion transition at 935 nm. (b) Relevant level schemes of a neutral InAs QD (left) and an \yb{174} ion (right). The $\ket{\textrm{0}}$--$\ket{\textrm{+1}}$ transition of the QD is on resonance with the \state{2}{D}{3/2}--\state{3}{D[3/2]}{1/2} transition of \yb{174}. At 4.2 T the $\ket{\textrm{0}}$--$\ket{\textrm{-1}}$ transition is detuned by 75 GHz and is not addressed. The S-P transition of \yb{174} at 369 nm is used for laser cooling and state readout of the ion. The number insets (in parentheses) are the cavity-modified (natural) \yb{174} branching ratios. (c) Absorption spectra of the QD (left) and the ion (right) transitions centered at 935 nm. The 60-fold mismatch in the radiative linewidths is further exacerbated to 93 including power broadening and spectral wandering effects.}
\end{figure*}
\indent Here, we report the first direct photonic coupling between a semiconductor QD spin and a trapped ion and demonstrate that single photons from a QD change the internal state of an $\textrm{Yb}^+$ ion efficiently, despite a significant mismatch in the optical properties of the two systems. To achieve this we link the atomic and solid-state nodes with single photons transmitted through an optical fiber [Fig. 1(a)].\\
\indent The atomic node consists of a single \yb{174} ion in a radio-frequency (RF) Paul trap located inside a recently developed fiber-based high-finesse Fabry-Perot cavity \cite{Steiner2013}. The miniature RF-Paul trap is made of two very fine tungsten needles at 100$\,\mu$m distance giving rise to trap frequencies in the range of 2$\pi \times \textrm{1-3 MHz}$. Ion fluorescence at 369$\,$nm is collected by an in-vacuo objective (NA $=0.27$) with 2\% collection efficiency and guided onto a photomultiplier tube with 14\% efficiency. The fiber cavity  \cite{Hunger2010} is made from two single mode fibers (125$\,\mu$m diameter) where a negative lens is machined on each tip (radii of curvature $-300\pm50$ $\,\mu$m). After the machining process, the fibers are coated with a high reflectivity dielectric coating (asymmetric coating, $T=10\,$ppm and 100$\,$ppm) resulting in a cavity finesse of $\mathcal{F} = 20\,000$. The length of the cavity is $170\pm10$ $\,\mu$m and the mode waist is about 6.1 $\,\mu$m. The ion interacts with a single mode of the optical cavity through the \state{3}{D[3/2]}{1/2}--\state{2}{D}{1/2} transition at 935 nm [Fig. 1(b)] in the intermediate coupling regime with cavity quantum electrodynamics (cQED) parameters  $(g,\kappa,\gamma)=2\pi\times (1.6,25,2.1)$ MHz. Here, $g$ denotes the coupling strength between the ion and the cavity mode, $\kappa$ the decay rate of the cavity field, and $\gamma$ the decay rate of the atomic dipole moment. Figure 1(c) (right) displays the corresponding absorption spectrum for this transition. The cavity-modified decay probability from the \state{3}{D[3/2]}{1/2} to the $D_{3/2}$ state is given by the bare probability of $2\%$ plus the Purcell factor of $2C_0/(2C_0+1)$ where we have used the cooperativity $C_0=g^2/(2\kappa \gamma)$, nominally 2.4(5)\% in the experiment.\\
\indent The solid-state node consists of an Indium Arsenide (InAs) QD in a Schottky diode placed inside a 4.2-K magneto-optical bath cryostat [Fig. 1(a)] giving access to neutral and negatively charged QD configurations, as well as electric and magnetic field tuning of the optical transitions  \cite{Warburton2000}. The samples contain a distributed Bragg reflector formed from alternating GaAs/AlGaAs  layers below the QD layer to increase the collection efficiency around 920-960$\,$nm. A superhemispherical zirconia solid immersion lens (Weierstrass geometry) mounted on the top surface of the sample is used to further increase the photon collection and improve spatial resolution. Resonant optical excitation and collection from single QDs is achieved using a confocal microscope with a 90:10 beam-splitter \cite{Matthiesen2012}. We collect both QD fluorescence and laser scattering via a 0.5 NA aspheric lens. Linearly polarized excitation, and cross-polarized detection allows us to suppress the laser scattering by a factor of $10^7$. For an excitation intensity of $I=I_{\mathrm{sat}}$, where $I_{\mathrm{sat}}$ denotes the measured laser intensity for which the steady-state QD excited state population is $1/4$, the signal-to-laser background ratio is 70:1 (20:1) in single-(two-)laser experiments.  Pulsed laser excitation is realized using acousto-optic modulators. Figure 1(c) (left) shows the absorption spectrum for the $\ket{0}\rightarrow \ket{+1}$ transition illustrated in Fig. 1(b). The inset displays the full emission spectrum consisting of the zero-phonon line and the spectrally broad phonon sideband. \\
\indent The resonantly generated QD photons are coupled into a 50-m long optical fiber and transmitted to the atomic node located 25 meters  away. The overall transmission probability of $5\times10^{-4}$ is given by the photon-extraction efficiency from the QD sample ($3.5\%$ into the first lens) and losses in the optical link from the QD to the ion ($1.4\%$ transmission from the first lens to the cavity mirror) \cite{supplementary}.\\
\begin{figure}
\includegraphics[width=1\columnwidth,angle=0]{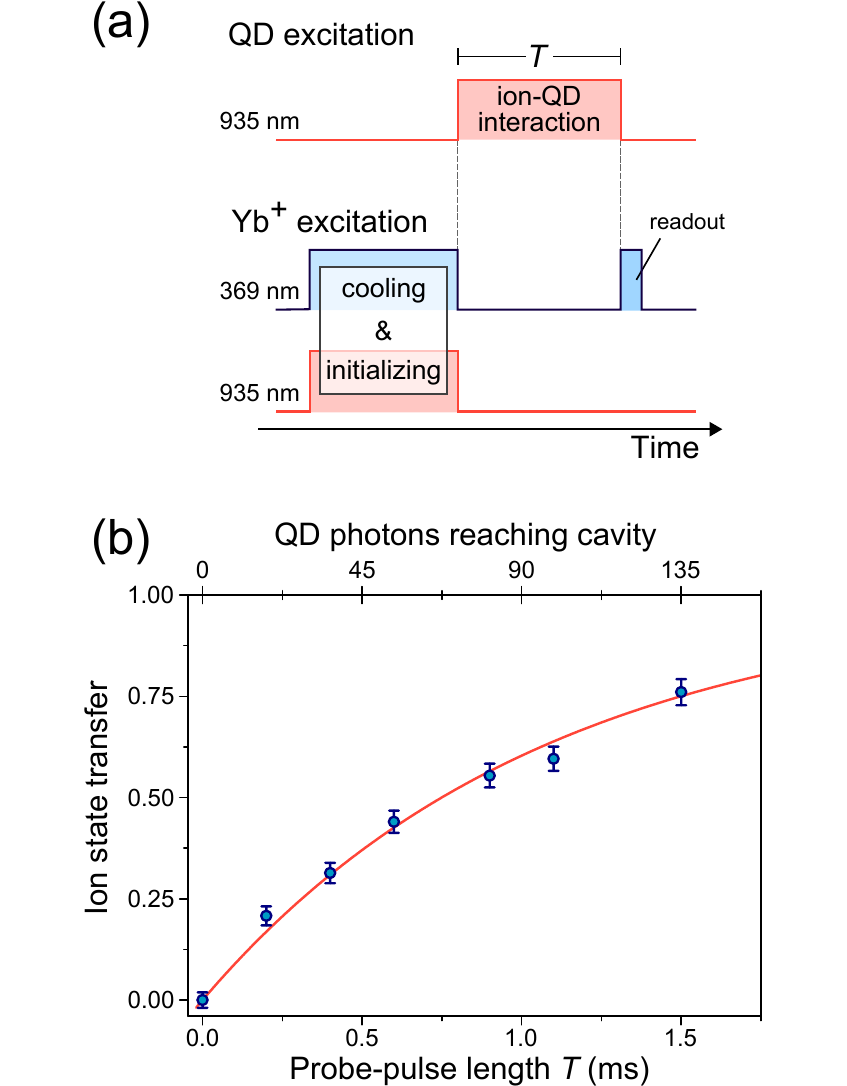}
\caption{\label{fig2}  (Color online) Optical pumping of the ion by QD photons. (a) The ion is prepared in the $\ket{-3/2}$ state  of the \state{2}{D}{3/2} manifold by optical pumping at 369 nm and 935 nm simultaneously. Subsequently, the QD is excited for a time $T$ and the generated photons are sent to the ion. The protocol cycle ends with the optical readout of the ion state. (b) Probability of the ion to absorb a QD photon during $T$ in the weak excitation regime ($I$=0.5$I_{\mathrm{sat}}$). The solid line is an exponential fit with a time constant $\tau=1.1$ ms. The top axis indicates the mean photon number for a given $T$ reaching the ion cavity and has 10\% statistical error.}
\end{figure}
\indent We first demonstrate the excitation of the atomic node with single photons from the solid-state node. We prepare the ion in the lowest Zeeman level $\ket{m_J=-3/2}$ of the \state{2}{D}{3/2} manifold by optical pumping. This state has a natural lifetime of 50\,ms and absorbs only $\sigma^+$--polarized photons. We then generate a single-photon stream from the bright neutral exciton transition ($\ket{0}\rightarrow \ket{+1}$) of the QD, as illustrated in Fig. 1(b). This transition has a radiative linewidth of $\Gamma_{\mathrm{QD}}=2\pi\times250 (10)$\, MHz, which is broadened further by spectral diffusion processes \cite{supplementary}. We drive the QD with an excitation intensity of $I=0.5I_{\mathrm{sat}}$ for a variable time $T$, which determines the total number of photons transmitted to the atomic node [see Fig. 2(a)]. Absorption of a QD photon transfers the ion into the \state{2}{S}{1/2} electronic ground state with a probability given by the intermediate \state{3}{D[3/2]}{1/2} state's cavity-modified branching ratio of 91:9. We then probe the \state{2}{S}{1/2}--\state{2}{P}{1/2} transition of the ion, where fluorescence at 369 \,nm verifies a successful state change. In contrast, we infer that a photon absorption event did not take place if the ion remains in the `dark' \state{2}{D}{3/2}-state \cite{supplementary}. Figure 2(b) displays the measured ion-state transfer probability as a function of $T$ for a fixed photon rate impinging on the fiber cavity of $\gamma_\textrm{QD}=9 (1)\times10^4\, \textrm{s}^{-1}$.  The exponential saturation behavior, displayed by the solid curve, yields a characteristic transfer time of $1.08 (4)$ ms, which corresponds to $97 (9)$ QD photons impinging on the cavity. This yields a single-photon absorption probability of $p_{\mathrm{abs}}=1.0 (2)\%$ at this excitation power. This is a conservative estimate as it includes the $13\%$ of the QD photons that are red-detuned by a few hundred GHz due to phonon-assisted emission \cite{Besombes2001, Matthiesen2013}, as seen in the inset of Fig. 1(c) and do not interact with the ion.\\
\begin{figure}[b]
\includegraphics[width=1\columnwidth,angle=0]{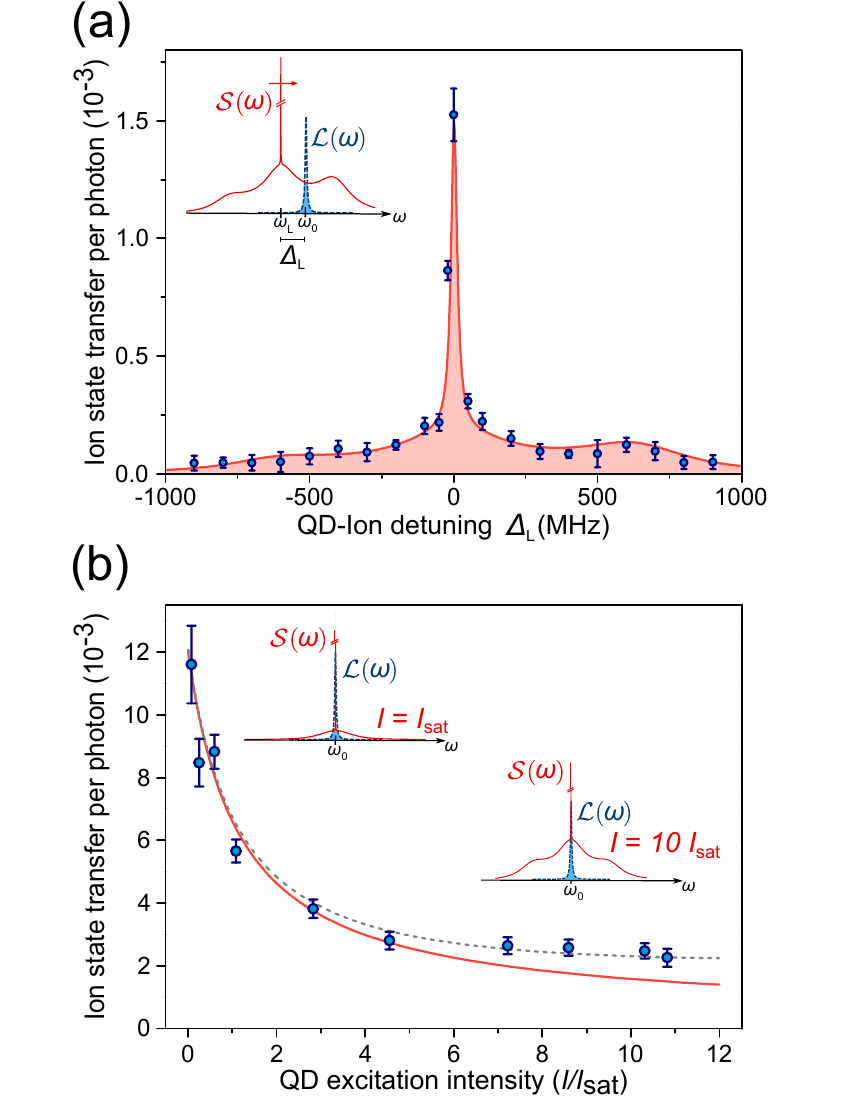}
\caption{\label{fig3} (Color online) Efficiency of ion state transfer as a function of the QD-ion detuning and QD driving intensity. (a) Spectral dependency of the absorption probability per received photon. The solid line displays the numerical model. The slight asymmetry is due to the detuning of the QD transition from the excitation laser frequency $\omega_{\mathrm{L}}$ induced by finite nuclear spin polarization, and the presence of fast dephasing at high excitation powers, characteristic for this sample \cite{supplementary}. (b) Absorption probability of the ion per received photon as a function of the QD excitation intensity. Insets show the QD photon spectrum $\mathcal{S}(\omega)$  at two different excitation intensities and the ion absorption line $\mathcal{L}(\omega)$. The convolution of the two, normalized by the number of QD photons, gives rise to the solid line. Including the imperfect suppression of the excitation laser (ratio of QD to laser photons is $70:1$ at $I_{\mathrm{sat}}$) in the numerical modeling predicts the dashed line. The error bars are statistical errors. We note that the ion-cavity coupling rate for the results in panel (b) is slightly higher than that for the measurements in panel (a).}
\end{figure}
\indent Figure 3(a) shows the dependence of the ion-state transfer probability on the spectral overlap between the QD photons and the ion transition. We tune the spectrum of QD emission in the strong excitation regime ($I=11I_{\mathrm{sat}}$) across the ion-cavity resonance and monitor the internal state of the ion \cite{supplementary}. The recorded state-transfer probability arises from the convolution of the QD single-photon spectrum, $\mathcal{S}(\omega)$, with the cavity-coupled ion absorption spectrum, $\mathcal{L}(\omega)$, as shown in the inset of Fig. 3(a), thus providing a measure of the spectral bandwidths of the two systems. As expected, the detuning dependence follows the Mollow-triplet signature of the QD emission spectrum. The narrow peak at zero detuning stems from the coherently scattered component as well as the residual laser due to imperfect suppression and its measured width, $\approx$ 20 MHz, is set directly by $\mathcal{L}(\omega)$. The solid curve gives the absorption spectrum calculated from the optical Bloch equations for the QD emission and ion-cavity absorption \cite{supplementary}. The highest probability of single-photon absorption for coherently scattered QD photons occurs at the exact cavity-ion resonance frequency, in stark contrast to the spectrally mismatched incoherent counterpart.\\
\indent Figure 3(b) displays the dependence of the ion-state transfer probability on the QD excitation laser intensity. We see a pronounced efficiency increase in the low excitation regime, where the dominant contribution to the QD photon spectrum is inherited from the continuous-wave (cw) excitation laser field  \cite{Matthiesen2013}. We measure a maximum value of $1.2(2)\%$ at $I=0.1I_{\textrm{sat}}$, which corresponds to 1.4(2)\% after accounting for the above-mentioned phonon-assisted emission. This value is comparable to the $1.8(2)\%$ measured independently for cw laser of equivalent intensity, which represents the upper limit on ion-photon interaction observed with this cavity. In the high-excitation regime ($I\gg I_{\textrm{sat}}$), where the main contribution to resonance fluorescence is incoherent, the absorption probability reduces by an order of magnitude, consistent with the theoretical prediction [solid curve in Fig. 3(b)]. The laser-like absorption probability demonstrates that coherent scattering can provide an efficient interface for systems presenting a strong radiative linewidth mismatch. Quantum-network protocols based on coherent scattering  \cite{Childress2005} are inherently probabilistic owing to the sub-unity photon generation rate in this regime. That said, our results still predict an overall 20\% higher efficiency of ion-state transfer over a deterministic generation scheme even when the coherent photon-generation probability is $10\%$ .\\
\begin{figure}
\includegraphics[width=1\columnwidth,angle=0]{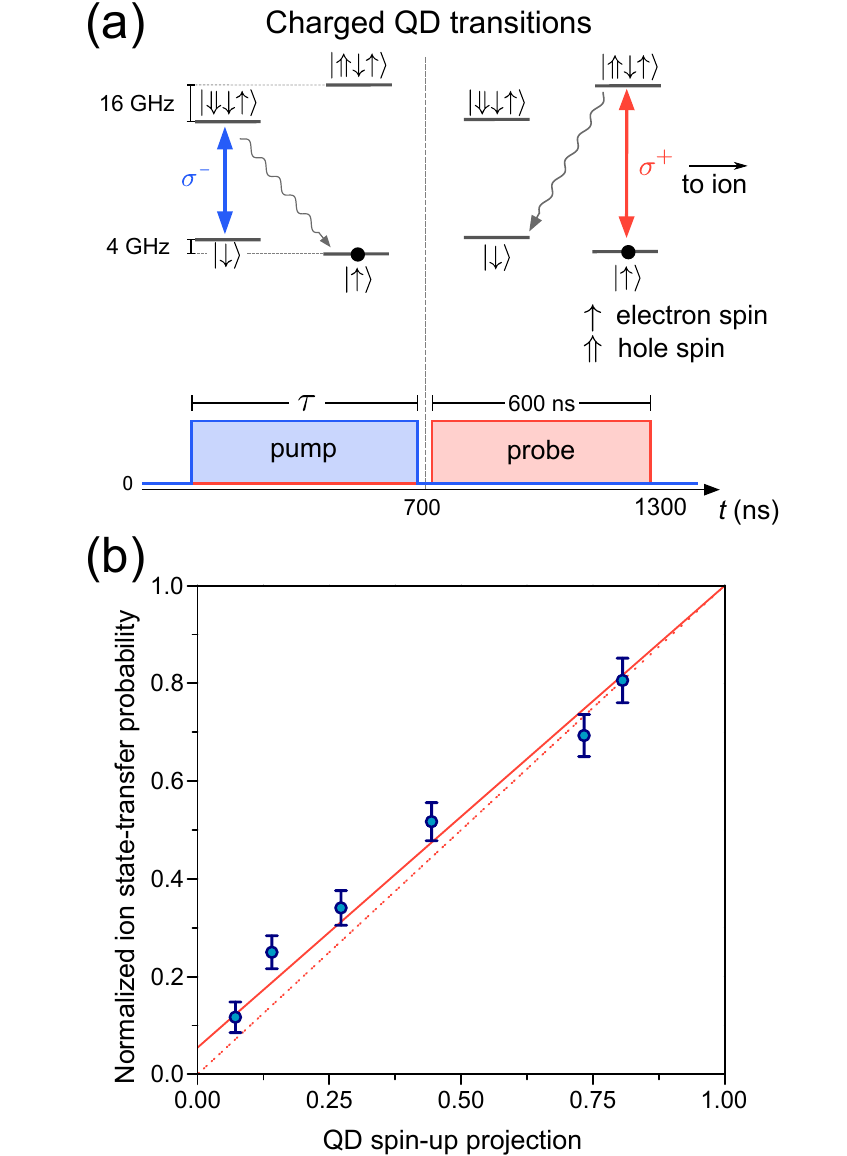}
\caption{\label{fig4}(Color online) Correlations between QD-spin projection and the state of the ion. (a) At $0.7$-T magnetic field the 20-GHz Zeeman splitting between the two transitions of a negatively charged QD ensures that only the $\ket{\uparrow}$-$\ket{\Uparrow\downarrow\uparrow}$ transition is resonant with the ion. The spin is prepared by optical pumping for a finite duration of $\tau$. Then, the $\ket{\uparrow}$-$\ket{\Uparrow\downarrow\uparrow}$ transition is driven for a fixed time of 600 ns and the generated photons are sent to the ion. (b) The measured spin-projection dependence of the normalized ion-state transfer (solid circles). The solid (dashed) curve is the expected dependence including (excluding) imperfect laser rejection.}
\end{figure}
\indent As a prerequisite for quantum-state transfer, we demonstrate classical communication between our solid-state and atomic nodes such that the internal state of the ion and the projection of the QD spin are correlated. First, we switch to a negatively charged QD under 0.7-T magnetic field in Faraday configuration. This provides optical access to the spin projection of the QD states via the Zeeman splitting of the ground and excited states [Fig. 4(a)]. The $\ket{\uparrow}$--$\ket{\Uparrow\downarrow\uparrow}$ transition is tuned  on resonance with the ion transition, while the $\ket{\downarrow}$--$\ket{\Downarrow\downarrow\uparrow}$ transition is off-resonant by $20$ GHz. Second, we prepare the desired spin mixture through optical pumping by driving the $\sigma^{-}$ transition with a pulse of variable duration, $\tau$, and the $\sigma^{+}$ transition with a 600-ns probe pulse. This alternating-pulses protocol provides a $\sigma_z$ projection of the electron spin ranging from $p_\uparrow=0.072(2)$ to $p_\uparrow=0.81(1)$ \cite{supplementary}. State-preparation and photon-generation steps are alternated with a repetition rate of $670\,$kHz during the QD-ion interaction time of 700\,$\mu$s. Once again, the ion is prepared in the $\ket{m_J=-3/2}$ Zeeman state of the $^2D_{3/2}$ manifold and absorbs $\sigma^+$--polarized photons leading to a state transfer to \state{2}{S}{1/2}. Figure 4(b) presents the theoretically expected as well as the measured correlation between the QD spin state and the internal state of the ion. The dashed curve indicates the ideal correlation for our state-transfer experiments, while the solid curve represents the expected correlation calibrated for the presence of residual laser background. This sequence maps the $\sigma_z$--spin component of the QD to the internal state of the ion within an average uncertainty of 3.8\%. Our results show that an arbitrary QD spin projection is reproduced faithfully on the atomic node in the form of \state{2}{S}{1/2} internal-state projection.\\ 
\indent The optical interface demonstrated here links two quantum systems with significantly different optical characteristics via the exchange of single photons. By coherent photon generation and cQED techniques we have achieved direct coupling between these systems with an efficiency that surpasses limitations set by their intrinsic properties. Our work can be extended to achieve faithful quantum-state transfer and distant entanglement between a QD and an ion. The hyperfine states of the trapped ion (e.g. in $^{171}$Yb$^{+}$) can serve as a long-term quantum memory for the QD spin qubits. A key challenge for the implementation of such a scheme is reaching sufficient coupling strength between the nodes.
While the overall efficiency of ~5$\times10^{-6}$ is shown here, we note that a 20-fold improvement in QD-photons collection efficiency has been recently achieved \cite{Dousse2010}; without in-situ monitoring of photons, a 10-fold reduction of loss in the optical link is straightforward, and increasing the absorption probability by a factor 30 through cavity improvements is within reach. Collectively, these technical steps could provide a three orders of magnitude improvement in our current node-to-node coupling in a not too distant future.\\
\indent We  acknowledge  support by the University of Cambridge, the Alexander-von-Humboldt Stiftung, EPSRC (EP/H005676/1), the European Research Council (Grant numbers 240335 and 617985), EU-FP7 Marie Curie Initial Training Networks COMIQ and S$^3$NANO. We thank J. Hansom, C. Schulte and J. Barnes for fruitful discussions and technical assistance. 

\section*{Supplementary Information}

\section{QD optical characterization}

\begin{figure}[htbp]
	\centering
		\includegraphics[width=1.00\columnwidth]{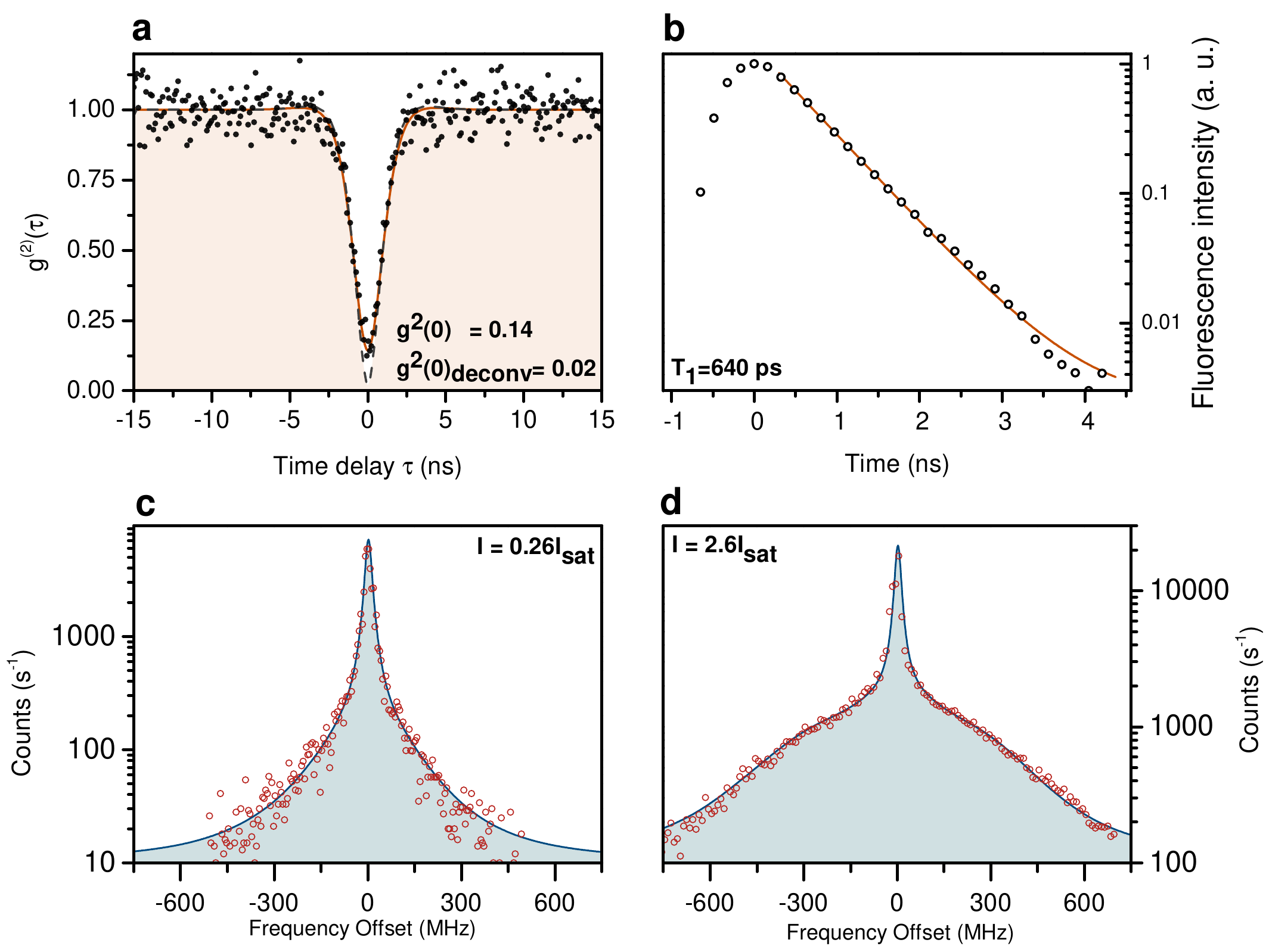}
	\caption{\label{QDcaracterisation}\textbf{QD characterization.} (a) Intensity correlation of the QD resonance fluorescence. The solid curve is a fit to the data, including spectral wandering of the transition and the response time of the detectors. The dashed curve is the autocorrelation that would be expected with an infinitely fast response. (b) Time-resolved fluorescence of the QD under pulsed excitation, showing a decay of 640 ps. (c) \& (d) Emission spectra of the QD for different excitation intensities, recorded with a scanning Fabry-Perot cavity. The solid lines are theoretically expected spectra, convolved with the 30 MHz width of the cavity. The theory includes pure dephasing, as detailed in section 4, and spectral wandering of the QD resonance of 300 MHz. The spectra are recorded at zero magnetic field, resulting in spectral fluctuations differing from those measured in the main experiment.}
\end{figure}

We present the measured optical properties of the neutral QD which was used in Fig. 1-3 of the main manuscript. Fig. \ref{QDcaracterisation}(a) is an intensity correlation measurement of the QD resonance fluorescence which demonstrates the single-photon nature of the QD emission. From this measurement and from the lifetime measurement presented in Fig. \ref{QDcaracterisation}(b), we can infer an excited-state lifetime $T_1=640$ ps, corresponding to a decay rate, $\Gamma$, of 2$\pi \times \textrm{250 MHz}$. We verify through Fabry-Perot scans the portion of coherent scattering from the QD, as shown in the spectra of Fig. \ref{QDcaracterisation}(c) and \ref{QDcaracterisation}(d). We also extract spectral wandering, and pure dephasing from modeling these spectra (described in section 4), resulting in the solid curves in Fig. \ref{QDcaracterisation}(c) and \ref{QDcaracterisation}(d).

\section{Determination of the absorption probability per photon $p_\text{abs}$}
\begin{figure}
\includegraphics[width=1\columnwidth,angle=0]{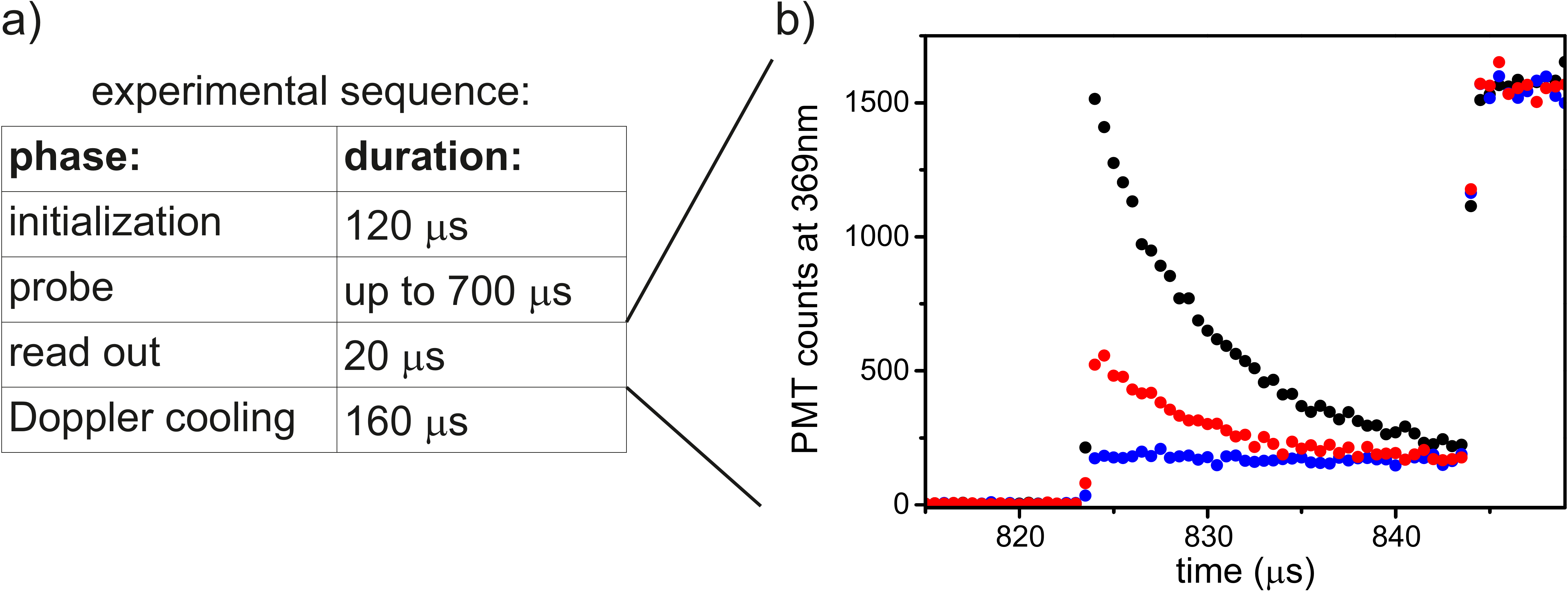}
\caption{\label{calibration}\textbf{$^{174}\text{Yb}^+$ initialization and readout.} (a) Experimental sequence. (b) Time-resolved 369\,nm fluorescence of the ion during the read out phase (starts at $t=824\,\mu$s). Black curve: ion is in the bright $^2$S$_{1/2}$ state. Blue curve: ion is in the dark $^2$D$_{3/2}$ state.  Red curve: the QD-photons transferred the ion partially from the dark to bright state. Each curve represents the accumulated data from 50000 repetitions.}
\end{figure}

The experimental sequence used to determine the absorption probability per photon $p_\text{abs}$ consists of four phases [cf. Fig. \ref{calibration}]. 

\begin{itemize}
	\item \textbf{Initialization}$\;$ The ion is prepared in the $m_J=-3/2$ Zeeman level of the $^2$D$_{3/2}$ state by applying $\sigma^-$ and $\pi$ polarized 935\,nm light in combination with 369\,nm light. The 935 nm beam has an angle of 75 degrees with respect to the cavity axis and the 369nm beam is transverse to it. After $120\,\mu$s, approximately 90\% of the population is accumulated in the target state.
	
	\item \textbf{Probe}$\;$ For a time $T_\text{interact} = 0-1500\,\mu$s the QD fluorescence is guided onto the fiber-cavity. The cavity length is actively stabilized to be resonant with the $^3$D[$3/2$]$_{1/2}$-$^2$D$_{3/2}$ transition of $^{174}\text{Yb}^+$ for all experiments presented in the main text. During $T_\text{interact}$ we record the number of photons reflected from the cavity. From this measurement we determine the average number of QD-photons $\bar{n}$ which impinged upon the cavity by taking account of the finite cavity in-coupling, attenuation on the optical path, background light and dark-counts of the photon detectors. The corresponding QD-photon rate is then $\gamma_\text{QD} =\bar{n}/T_\text{interact}$.   
	
	\item \textbf{Readout}$\;$ In order to prove that the ion has been interacting with a QD photon during $T_\text{interact}$ we make use of the large branching ratio from the $^3$D[$3/2$]$_{1/2}$ to the $^2$S$_{1/2}$ state. Since the $^3$D[$3/2$]$_{1/2}$ state decays mainly to the $^2$S$_{1/2}$ state we define $p_\text{abs}$ as the probability per photon to transfer population from $^2$D$_{3/2}$ to $^2$S$_{1/2}$. The effect of the QD photons is then determined by subsequent scattering of 369\,nm light on the $^2$P$_{1/2}$-$^2$S$_{1/2}$ transition of the ion. 
	
	Fig. \ref{calibration}(b) shows the 369\,nm fluorescence detected by a photo-multiplier tube (PMT) during the read-out phase. The fluorescence signal decays exponentially if the ion is in $^2$S$_{1/2}$ state (black curve). This  decay occurs because the 369\,nm transition is not cyclic. In contrast, for an ion in the $^2$D$_{3/2}$ state the read out signal is constant; the finite signal amplitude arises from background light (blue curve). The red curve shows the signal for a stream of QD photons. The $^2$D$_{3/2}$ to $^2$S$_{1/2}$ transfer probability $p_\text{transfer}$ for the QD probe pulse is calculated by integrating the PMT counts during the first $19\,\mu$s of the readout phase for the three traces, giving $c_\text{D-state}$,$c_\text{S-state}$ and $c_\text{QD}$:
\begin{equation}
p_\text{transfer}:=\frac{c_\text{QD} -c_\text{D-state} }{c_\text{S-state}-c_\text{D-state}}\,.
\end{equation}
$p_\text{abs}$ is determined from $p_\text{transfer}$ and the mean number of photons $\bar{n}$:
\begin{equation}
p_\text{abs}:=  \frac{\Gamma_\text{abs}}{\gamma_\text{QD}} = \frac{-\ln(1-p_\text{transfer})}{T_\text{interact}} \frac{T_\text{interact}}{\bar{n}} = \frac{-\ln(1-p_\text{transfer})}{\bar{n}}\,.
\end{equation}
As not every excitation to $^3$D[$3/2$]$_{1/2}$ state leads to decay to the $^2$S$_{1/2}$ state $p_\text{abs}$ underestimates the actual excitation probability. From the cavity-QED parameter we infer that the altered branching ratio is 91\% in favour of the decay to $^2$S$_{1/2}$ state and therefore our measurements systematically underrate $p_\text{abs}$ by 9\%. 
	
	\item \textbf{Doppler cooling}$\;$ Finally, the ion is continuously laser-cooled on the  $^2$P$_{1/2}$-$^2$S$_{1/2}$ transition for $160\,\mu$s.
\end{itemize}

\section{Modeling of the photonic link}
The spectrum of the QD emission $S(\nu)$ is given by the Fourier transform of the two-time correlation function of the optical dipole $\left\langle\sigma_+(\tau)\sigma_-(0)\right\rangle$.
To calculate the QD spectrum, we consider a two-level system interacting with a classical field in the rotating wave approximation (e.g. \cite{Scully}) with an additional pure dephasing term $\gamma$ of the optical dipole. The relevant physical quantities are:

\begin{itemize}  
  \item The excited state decay rate $\Gamma= 2\pi \times 250$ MHz.
  \item The QD-laser Rabi coupling $\Omega$, or equivalently, the normalized intensity $I/I_{sat}\propto\Omega^2/\Gamma^2$.
  \item An intensity-dependent dephasing rate of the optical dipole $\gamma= 2\pi \times (I/I_{sat})\times9.3$ MHz.
  \item The frequency of the laser $\nu_L$.
  \item The detuning between the laser and the optical dipole frequency $\delta=\nu_{QD}-\nu_L$.
  \item The frequency of the ion $\nu_0$.
\end{itemize}

The number of QD photons $n_{QD\rightarrow Ion}(\nu_{0})$ spectrally overlapping with the ion resonance is:
\begin{eqnarray}
n_{QD\rightarrow Ion}(\nu_{0})=\int S(\nu)L(\nu_0-\nu)d\nu \, ,
\end{eqnarray}
where $L(\nu_0-\nu)$ is a Lorentzian of FWHM of $20$ MHz corresponding to the width of the cavity-mediated ion absorption.

The transfer rate from $^2$D$_{3/2}$ to $^2$S$_{1/2}$ ($\Gamma_{abs}$) is assumed to be proportional to $n_{QD\rightarrow Ion}(\nu_{0})$, with a correction for a small portion of laser photons $n_L$ which leak to the ion due to finite polarization rejection:
\begin{eqnarray}
\Gamma_{abs}\propto n_{QD\rightarrow Ion}(\nu_{0})+n_L*L(\nu_0-\nu_L)\, .
\end{eqnarray}
And the absorption probability per photon is proportional to:
\begin{eqnarray}
p_{abs}^{model}\propto \frac{n_{QD\rightarrow Ion}(\nu_{0})+n_L*L(\nu_0-\nu_L)}{n_{QD}+n_L} \, ,
\end{eqnarray}
where $n_{QD}$ is the total number of QD photons. The modeled $p_{abs}^{model}$ is scaled to fit the observation. The scaling factor is governed by the ion-cavity coupling.

As stated in the methods, the QD signal to laser leakage is $70:1$ at saturation. Consequently, in the modeling, we take:

 \begin{eqnarray}
n_L(I_{sat})&=&n_{QD}(I_{sat})/70 \, \\
n_L(I) &\propto& I \, .
\end{eqnarray}

\begin{figure}
\includegraphics[width=1\columnwidth,angle=0]{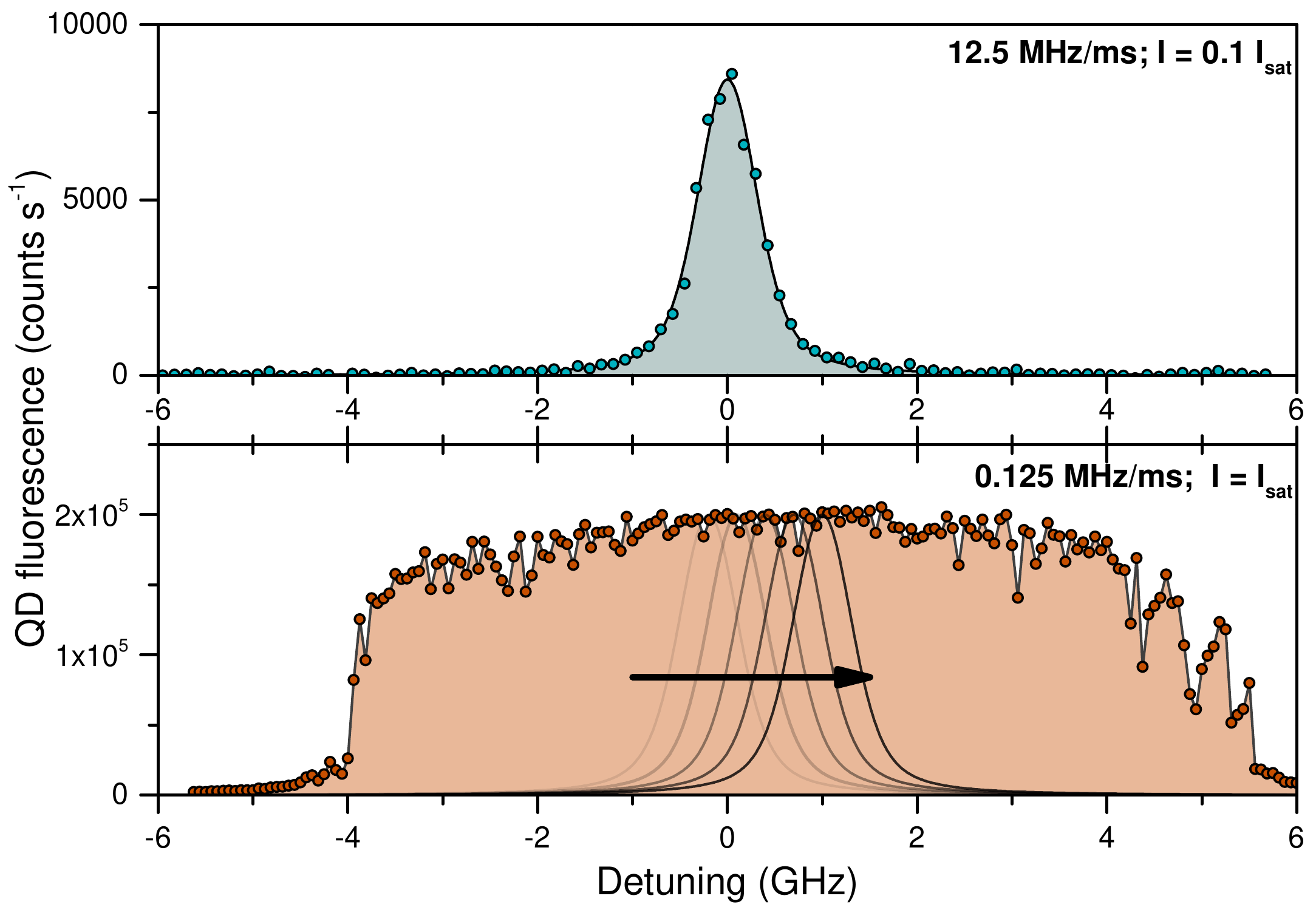}
\caption{\label{dragging} \textbf{QD absorption scans.} Absorption scans of the neutral QD used in the main manuscript. If the laser scanning rate is sufficiently slow, strong deviation from a Lorentzian lineshape can be observed as a hyperfine field develops to ensure through an effective Zeeman shift that the QD transition remains nearly resonant with the laser frequency. The Lorentzian curves shown in the bottom panel illustrate how the absorption profile is continuously shifted to follow the laser.}
\end{figure}

In order to model the spectrum, shown in Fig. 3(a) of the text, we computed $p_{abs}^{model}$, as a function of $\nu_0-\nu_L$. Having measured the lifetime of the QD, the QD dipole dephasing rate and the FWHM of the ion absorption through the cavity independently, the detuning is the only free parameter. The asymmetry observed experimentally in Fig. 3(a), main text, results from the combined effect of optically induced dephasing and detuning \cite{Ulhaq2013}.
The theoretical curve in Fig. 3(a) of the main text illustrates such an imbalance with $\delta=250\,$MHz and $\gamma= 2\pi \times 93\,$MHz. Detuning is set by a non-linear feedback mechanism known as dragging: the QD transition is locked to the resonant laser, and the locking point is determined by hyperfine interaction between the optically created electron and the nuclei in the QD \cite{Latta2009, Vink2009, Hogele2012}. Typical absorption scans, taken for two different scanning speeds and excitation intensities, are presented in Fig. \ref{dragging}. As shown in the figure, higher excitation intensities and slower scanning speeds permit a build up of nuclear polarization, holding the QD resonance to the excitation frequency. The emission profile is exactly the same as in the absence of dragging. This dragging of the QD transition is used for tuning the emission spectrum presented in Fig. 3(a) of the main manuscript.

$p_{abs}^{model}$ is computed for $\nu_0=\nu_{QD}=\nu_L$ as a function of the normalized excitation intensity $I/I_{sat}$ to model the intensity-dependent absorption shown in Fig. 3(b) of the main text. The modeled evolution is scaled to fit the data. This scaling factor is governed by the ion-cavity coupling. An excitation-dependent pure dephasing term has been included, for consistency with the spectrum modelling in Fig. 3(a) of the main text, but its effect is below our experimental resolution.

For semiconductor QDs, relaxation between the dressed states induced by acoustic phonons sets a lower bound to the dephasing rate observed at a given Rabi frequency \cite{Ramsay2010,Ulhaq2013}. For a saturation parameter $I/I_{sat}=10$, a dephasing rate on the order of $0.1-1\,$MHz is to be expected from such processes. For this QD, a much larger intensity dependent dephasing rate was measured which is likely to be caused by fast electrical field noise in this sample. Excitation intensity-dependent spectral wandering (i.e. slow electrical field fluctuations) with a $160$ MHz amplitude at $I/I_{sat}=1$ and a $250$ MHz amplitude at $I/I_{sat}=5$ as well as on-off switching of the fluorescence were also observed. The effect of spectral wandering in the modeling was found to be a small correction compared to the accuracy of the data points, so it was not included in our final analysis. 

\section{Photon Transfer Efficiency}

In addition to the ion state transfer per photon probabilities reported there is a constant transfer efficiency from the QD to the ion cavity of $5\times10^{-4}$, as mentioned in the main text. This can be partitioned into two main stages: photon out-coupling from the QD sample (3.5\%) and losses in the path from the sample to the ion cavity (1.4\% transmission).  A large proportion of the losses between the QD sample and the ion cavity can be attributed to the need to monitor photon rates between the systems, as well as polarization control between the QD and the ion. 


The quoted transmission can be recovered from the individual elements which make up the link from the QD sample to the ion cavity: QD microscope beam splitters ($90\%\times2$ transmission), linear polarizer (41\%), coupling into the QD microscope fiber (40\%), coupling into the 50-m fiber (70\%), polarization optics transmission at 50-m fiber output (90\%), polarization filtering (50\%), beam splitter transmission ($90\%\times2$) and coupling into the fiber cavity (42\%).

\section{Preparation of the QD electron spin mixture}

\begin{figure}
\centering

\includegraphics[width=\columnwidth,angle=0]{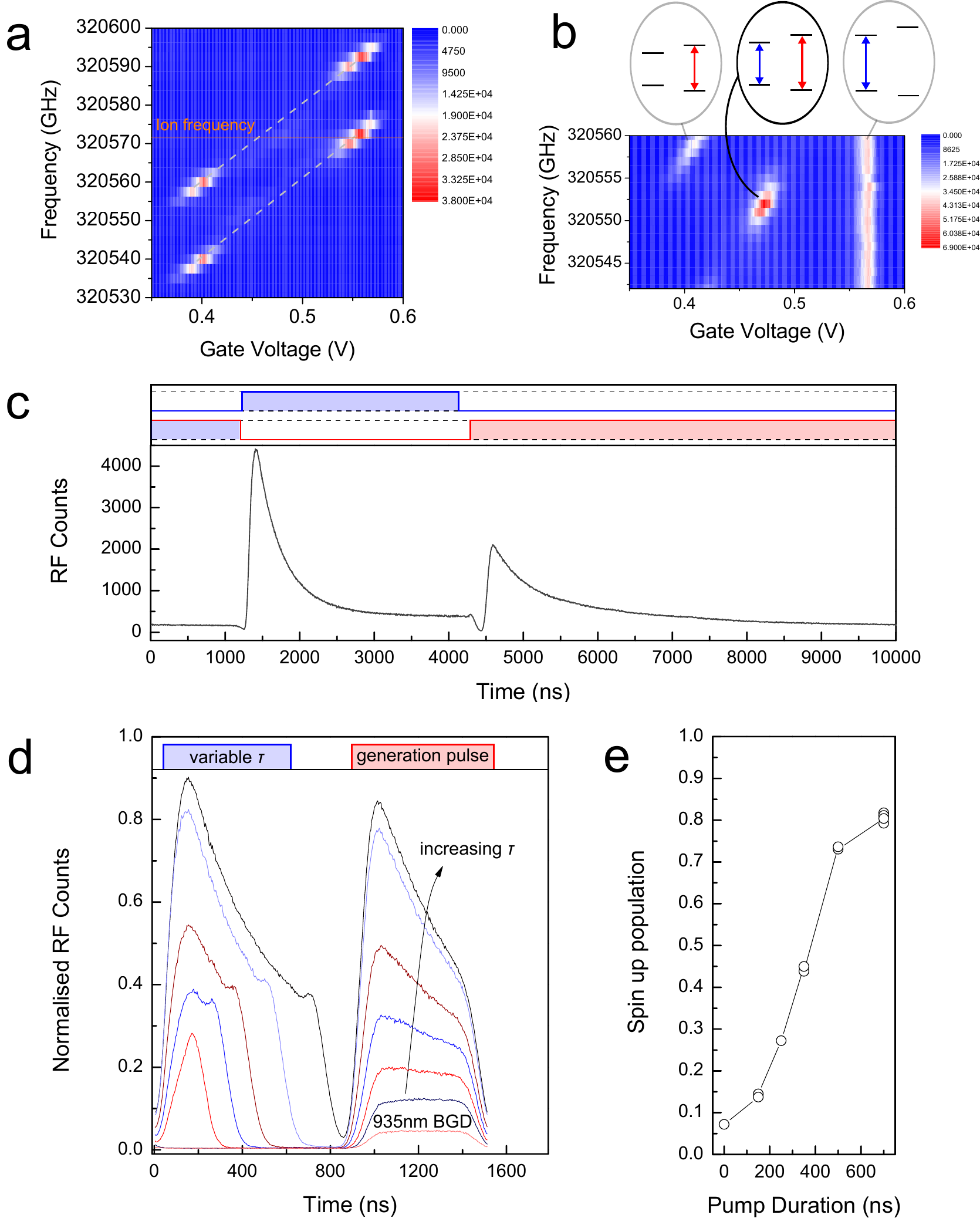}
\caption{\label{SpinPumping}\textbf{QD spin preparation.}(a) Single laser scan: 2D map of the QD fluorescence intensity as a function of laser frequency and gate voltage. An external magnetic field of $\mathrm{B}=0.7\,$T is applied in Faraday geometry. (b) Two-laser scan: 2D map of the fluorescence intensity as a  function of the frequency of a scanning laser (red arrow in the diagrams) and gate voltage while a second laser (blue arrow) is set at a fixed frequency of $320572\,$GHz (corresponding to the ion frequency). The strong resonance observed around a gate voltage of $0.475\,$V and $320552\,$GHz occurs when each laser is resonant with one of the negatively charged QD transitions. The other high fluorescence regions are due to single laser resonances in the co-tunneling as observed in (a). (c) Time-resolved fluorescence under alternating $320572\,$GHz/$320552\,$GHz excitation. Laser background has been subtracted. (d) Time-resolved resonance fluorescence (not background subtracted) under alternating $320552\,$GHz/$320572\,$GHz excitation in the same experimental conditions as for spin-state dependent absorption from the ion, and (e) corresponding electron spin population.}
\end{figure}

In order to demonstrate a dependence of the ion absorption on the spin state of a QD, we split the transitions of a negatively charged QD with an external magnetic field applied along the growth axis ($B=0.7\,$T). While polarization-spin correlations are erased by our cross-polarization technique, energy-spin correspondence is used to provide a spin-selective absorption. 

The energy of the transitions is found by mapping the intensity of the resonance fluorescence as a function of laser frequency and gate voltage as shown in Fig. \ref{SpinPumping}(a). Strong fluorescence is observed when the gate voltage is set to $0.4$ V ($0.55$ V) corresponding to fast co-tunneling as the electron occupation changes from 0 to 1 (1 to 2) in the QD. Fast co-tunneling randomizes the spin-state of the electron, preventing spin shelving following optical pumping. Between these two co-tunneling regions, a strong reduction of fluorescence occurs due to optical pumping. The fluorescence can be recovered using a second repump laser as shown in Fig. \ref{SpinPumping}(b) \cite{Atature2006}.

To fully characterize the optical pumping, we perform a time-resolved pulsed two-color experiment [Fig. \ref{SpinPumping}(c)]. Two pulses are generated from cw laser sources using acousto-optic modulators. A $3$ $\mu$s pulse at $320552$ GHz resonantly drives the red transition at a power corresponding to $I/I_{sat}=2$. The emission rate of red photons decreases exponentially in time as the electron is pumped into the $\ket{\uparrow}$ state. Then, a $6.5$ $\mu s$ pulse at $320571$ GHz resonantly drives the blue transition at $I/I_{sat}=0.5$. Again, the emission rate decreases exponentially as a consequence of optical pumping, this time into the $\ket{\downarrow}$ state. From these, we can estimate a preparation fidelity of $92.2\%\,\pm0.2\%$  ($92.8\%\pm0.2\%$) in the state $\ket{\uparrow}$ ( $\ket{\downarrow}$).

In the experiment showing spin-selective absorption of the ion, we use pulses which only differ from the previous experiment by their duration: the pulse resonant with the red transition has a variable duration between $0$ and $700$ ns, which is used to create an incoherent spin superposition  $p_{\uparrow} \ket{\uparrow} \bra{\uparrow} +p_{\downarrow}\ket{\downarrow}\bra{\downarrow}$ with $p_{\uparrow}$ varying between $0.072(2)$ and $0.81(1)$. Photons are then generated, mostly coherently, by a $550$ ns pulse resonant with the blue transition. Time-resolved resonance fluorescence traces corresponding to identical experimental conditions are presented in Fig. \ref{SpinPumping}(d). In this protocol, the 935\,nm generation pulse is kept constant, so that the change of ion absorption can only be attributed to a change of spin preparation of the QD electron.

\end{document}